# Title

Josephson scanning tunneling microscopy – a local and direct probe of the superconducting order parameter

**Authors**

Hikari Kimura[1,2,†], R. P. Barber, Jr.[3], S. Ono[4], Yoichi Ando[5], and R. C. Dynes[1,2,6*]

[1]*Department of Physics, University of California, Berkeley, California 94720, USA*

[2]*Materials Sciences Division, Lawrence Berkeley National Laboratory, Berkeley, California 94720, USA*

[3]*Department of Physics, Santa Clara University, Santa Clara, California 95053, USA*

[4]*Central Research Institute of Electric Power Industry, Komae, Tokyo 201-8511, Japan*

[5]*Institute of Scientific and Industrial Research, Osaka University, Ibaraki, Osaka 567-0047, Japan*

[6]*Department of Physics, University of California, San Diego, La Jolla, California 92093-0319*

*rdynes@ucsd.edu

†Present address: *Department of Physics and Astronomy, University of California, Irvine, California 92697, USA*

**Abstract**

Direct measurements of the superconducting superfluid on the surface of vacuum-cleaved $Bi_2Sr_2CaCu_2O_{8+\delta}$ (BSCCO) samples are reported. These measurements are accomplished via Josephson tunneling into the sample using a scanning tunneling microscope (STM) equipped with a superconducting tip. The spatial resolution of the STM of lateral distances less than the superconducting coherence length allows it to reveal local inhomogeneities in the pair wavefunction of the BSCCO. Instrument performance is demonstrated first with Josephson measurements of Pb films followed by the layered superconductor $NbSe_2$. The relevant measurement parameter, the Josephson $I_CR_N$ product, is discussed within the context of both BCS superconductors and the high transition temperature superconductors. The local relationship between the $I_CR_N$ product and the quasiparticle density of states (DOS) gap are presented within the context of phase diagrams for BSCCO. Excessive current densities can be produced with these measurements and have been found to alter the local DOS in the BSCCO. Systematic studies of this effect were performed to determine the practical measurement limits for these experiments. Alternative methods for preparation of the BSCCO surface are also discussed.

## I. INTRODUCTION

Scanning tunneling microscopy (STM) and scanning tunneling spectroscopy (STS) have been extensively utilized for nanometer scale studies of physical and electronic structures on the surface of high transition temperature (high-$T_C$) superconducting cuprates, especially $Bi_2Sr_2CaCu_2O_{8+\delta}$ (BSCCO) and $YBa_2Cu_3O_{7-\delta}$ (YBCO). A rich array of experiments includes imaging vortex flux lines[1,2,3] and mapping the integrated local density of states (LDOS).[4,5,6] These latter results were used to determine a "gap" $2\Delta$ taken to be the difference between two coherence peaks in the LDOS. The spatial resolution combined with spectroscopic results has been used to produce "gap maps" over the surface for various dopings in BSCCO. These studies have revealed that (i) the formation of gapped regions obtained from the $dI/dV$ spectra actually start above $T_C$, and there is a linear relation between $\Delta$ and the gap opening temperature, $T^*$;[7]

and (ii) there is an anticorrelation between the energy gap, $\Delta$ in the superconducting state and the normal state conductance at the Fermi energy[8] for various dopings of BSCCO. A challenge with this approach is that the identification of $\Delta$ becomes ambiguous in strongly underdoped BSCCO where the observed $dI/dV$ curves no longer have well-defined sharp coherence peaks.[9,10]

Although these STM/STS experiments on high-$T_C$ superconductors have yielded a wealth of data, they suffer from important limitations. Of particular significance in our opinion is that because they utilize a normal metal tip, these studies can only probe the local *quasiparticle* density of states (DOS). This DOS almost certainly has an intimate connection with the superconductivity in these materials; however that relationship is still unknown. Furthermore, the derived "gap" qualitatively changes its shape with doping. In contrast, the BCS theory for conventional superconductors defines a well-established relationship between the gap in the quasiparticle DOS ($\Delta_{BCS}$) and fundamental quantities of the superconducting state including the amplitude of the order parameter and the superconducting transition temperature, $T_C$. Without a similar theory for the high-$T_C$ superconductors to enable the inference of the superconducting properties from the quasiparticle DOS, it is necessary to directly probe the superconducting superfluid of these materials. Two central questions that such a direct probe should address are (i) whether the superconducting order parameter of BSCCO has spatial variation, and (ii) how the superconducting ground state correlates with the quasiparticle excited states ($\Delta$).

In this article we will present the results of experiments using an STM with a superconducting tip. This instrument allows us to directly probe the superconducting superfluid at the surface of BSCCO using the Josephson effect. We will first discuss the primary quantity derived from Josephson tunneling measurements, the $I_C R_N$ product, and its relationship to the fundamental properties of the superconductors that make up the tunnel junction. After a description of the technical aspects of the apparatus, data which verify its successful operation on a conventional superconductor (Pb) will be presented. The approach is then extended to a layered superconductor (NbSe$_2$), and then finally to the high-$T_C$ superconductor BSCCO. BSCCO is believed to be primarily a *d*-wave superconductor but with a weak *s*-wave component. Furthermore, the STM probes the order parameter at the surface where the symmetry restrictions might be relaxed somewhat. This experiment relies on coupling to the *s*-wave component.

## II. $I_C R_N$ PRODUCT

Josephson tunneling is Cooper pair tunneling between two superconductors separated by a thin barrier. The zero-voltage supercurrent flowing through the junction is given by $I = I_C \sin(\varphi_2 - \varphi_1)$, where $I_C$ is the maximum zero temperature supercurrent that the junction can sustain and $\varphi_{1(2)}$ is the phase of two superconducting electrodes' order parameter. The maximum supercurrent is related to the amplitude of the superconducting pair wave function. An STM with a superconducting tip can be a local Josephson probe and can, in principle, access the superconducting pair wave function directly on a length scale smaller than or comparable to the superconducting coherence length, $\xi$.

Between the two superconductors in a tunnel junction, the Josephson $I_C R_N$ product ($R_N$ is the normal state resistance of the junction) is a directly measurable quantity uniquely determined by the specific materials. $I_C R_N$ is a fundamental parameter that is directly linked to the superconducting order parameter amplitude $|\Psi|$, and in the case of conventional superconductors to the energy gaps $\Delta_{BCS}$ through the BCS relationship.[11] Josephson studies using a

superconducting STM on conventional superconductors have shown good agreement between the measured $I_C R_N$ and BCS predictions.[12,13] For high-$T_C$ superconductors, on the other hand, there is no established theory to relate $I_C R_N$ with $\Delta$ derived from the quasiparticle excitation spectrum. There can be at least two additional issues not present for conventional superconductors. The first is that we do not have a universally agreed-upon theory of the high-$T_C$ cuprates and secondly, the symmetry of the order parameter has a substantial $d_{x^2-y^2}$ component thus making the coupling to a conventional superconductor (*s*-wave symmetry) more complicated. An $I_C R_N$ measurement on BSCCO using a conventional superconducting STM should, however, both prove the existence and yield the amplitude of the BSCCO pair wave function that couples to the conventional superconducting tip. Because of the spatial resolution of an STM, this measurement could reveal useful new information regarding inhomogeneities in the superconductivity of BSCCO.

## III. EXPERIMENTAL

### A. Superconducting Tip

A reproducible and stable superconducting tip fabrication method that we have developed begins with a $Pt_{0.8}/Ir_{0.2}$ tip mechanically cut from a 0.25 mm diameter wire.[14] Tips are then placed in a bell-jar evaporator with the tip apex pointing towards the evaporation sources. 5500 Å of Pb is deposited at a rate of ~ 40 Å/s followed by 36 Å of Ag at a rate of 1 Å/s without breaking vacuum. The thick layer of Pb was chosen such that at 2.1 K, well below transition temperature of Pb ($T_C$ = 7.2 K), there would be bulk superconductivity in the tip (superconducting coherence length, $\xi_0$, of Pb is $\xi_0$ = 830 Å). The Ag serves as a capping layer to protect the Pb layer from rapid oxidation upon exposure to the atmosphere. The Ag layer is thin enough to proximity-couple to the Pb layer, resulting in a superconducting tip with $T_C$ and $\Delta_{BCS}$ only slightly below that of bulk Pb. Because the Pb/Ag bilayer is deposited without breaking vacuum, the interface between the layers is expected to be clean, and thus superconductivity may be induced in the Ag layer by the proximity effect.[15] The Pb/Ag combination is also a good metallurgical choice, as there is no significant alloying at the interface.[16] For Josephson measurements of a conventional superconductor, Pb/Ag samples were also evaporated onto freshly cleaved graphite substrates during the same deposition as the tips. The same Ag capping layer keeps these samples stable for the transfer from the evaporator to the STM apparatus.

### B. NbSe$_2$

Single crystal 2$H$-NbSe$_2$ was chosen as a first target material beyond conventional superconductors as a surrogate to the high-$T_C$ superconducting cuprates. 2$H$-NbSe$_2$, a family of layered transition-metal dichalcogenides, is a type-II conventional superconductor with $T_C$ = 7.2 K and charge density wave (CDW) transition at $T_{CDW}$ = 33 K, so the CDW state coexists with the superconducting state below $T_C$. This material also has short coherence lengths ($\xi_{a,b}$ = 77 Å and $\xi_c$ = 23 Å), an anisotropic $s$-wave gap varying from 0.7 to 1.4 meV across the Fermi Surface[17] and multiband superconductivity indicated from observations of momentum-dependent superconducting gap on the different Fermi surface sheets.[18] Van der Waals bonding between Se layers is so weak that the crystal is easily cleaved to expose a fresh and inert surface for STM measurements. However no Josephson tunneling measurements have been reported before those of our group,[13] partly because it is difficult to grow a stable insulating (usually oxide) layer for planar tunnel junctions.

## C. $Bi_2Sr_2CaCu_2O_{8+\delta}$

There is still much discussion and controversy about the symmetry of the order parameter of high-$T_C$ superconducting cuprates and whether the pseudogap state observed in the underdoped region in BSCCO is a precursor to the coherent superconducting state. If the symmetry is strictly *d*-wave, there should exist no Josephson coupling between high-$T_C$ superconducting cuprates and a conventional superconducting tip. However Josephson coupling has been observed between conventional superconductors and YBCO[19] as well as BSCCO.[20,21] Thus Josephson measurements of BSCCO using a superconducting STM should reveal new results about the symmetry of the BSCCO order parameter. Correlations between $I_CR_N$ products from the Josephson effect and the energy gap $\Delta$ measured from the quasiparticle excitation spectra of BSCCO should contribute to the construction of a microscopic theory of the mechanism of high-$T_C$ superconducting cuprates. An apparent constraint indicated by a previous study of BSCCO at high STM currents[5] that we will discuss later is that there is a threshold current above which the BSCCO morphology and spectroscopy are drastically and irreversibly changed. This effect appears at a tunneling current of about 500 pA. Care will be taken to sweep the bias of the STM junction such that the resulting tunneling current does not exceed this threshold to avoid changing the BSCCO electronic structure.

BSCCO single crystals are grown by the floating zone method[22] with the hole doping, $\delta_h$, ranging between heavily underdoped ($T_C$ = 64 K) and overdoped ($T_C$ = 74 K) via optimally-doped samples ($T_C$ = 94 K). $T_C$ was determined by magnetic susceptibility measurements. In these current studies, extensive Josephson measurements were performed on overdoped samples with different dopings ($T_C$ = 76 K, 79 K and 81 K). Our BSCCO samples which have typical dimensions of 1 mm × 1 mm and a few tens of μm thick are glued onto a copper plate with silver epoxy (Epoxy technology EE 129-4). All of the samples were cleaved in high vacuum (~ $3 \times 10^{-8}$ Torr) at room temperature to expose a fresh surface. BiO-BiO planes are attracted by weak Van der Waals bonding so that they are most likely the cleavage plane,[23] with the conducting $CuO_2$ surface two layers below. Cleaved samples are then cooled to $T = 2.1$ K after being inserted into the STM. The experiments are performed at 2.1K where the properties of Pb approach the low temperature limit. While lower temperatures would nominally reduce the fluctuations, we find that a characteristic noise temperature $T_n$ of the apparatus dominates the fluctuations. A lower base temperature in the current configuration would therefore not improve the measurements; only a significant reduction of the noise leakage from room temperature electronics would do so.

### D. $E_J$, phase fluctuations, $T_n$, current density

The signature Josephson response of a superconducting STM differs from that of typical low $R_N$ planar S/I/S devices. For identical superconductors, Ambegaokar and Baratoff[11] derived the temperature dependent Josephson binding energy, $E_J$, where

$$E_J(T) = \frac{\hbar I_C}{2e} = \frac{\pi \hbar}{4e^2} \frac{\Delta(T)}{R_N} \tanh\left(\frac{\Delta(T)}{2k_B T}\right). \qquad (1)$$

Because of the experimental base temperature ($T = 2.1$ K) and large $R_N$ associated with an STM, $E_J$ is smaller than $k_B T$ for the Josephson STM. For example, with an STM resistance of 50 k$\Omega$, $E_J/k_B$ is roughly 1 K. Also for ultra-small tunnel junctions, the Coulomb charging energy, $E_C$ can be large. We estimate the capacitance, $C$, of the STM junction formed between the conical tip apex and the sample surface to be about 1 fF. $E_C = e^2/2C$ is therefore of order 1 K: comparable to both $E_J$, and $k_BT$. The time-scale of an electron tunneling in the STM junction and conducting off the tip is much shorter than $\hbar/E_C$, so that the electron is swept away long before the charging effects become relevant. Because $k_BT$ is the dominant energy, the phase difference of the two superconductors, $\varphi$, is not locked in a minimum of the sinusoidal $E_J$ vs. $\varphi$ washboard potential, but is thermally excited and diffusive (Fig. 1). With a current bias, the phase diffuses preferentially in one direction as illustrated in Fig 1. Near zero bias voltage the observed Josephson current is therefore dependent on the bias due to the dissipative phase motion.

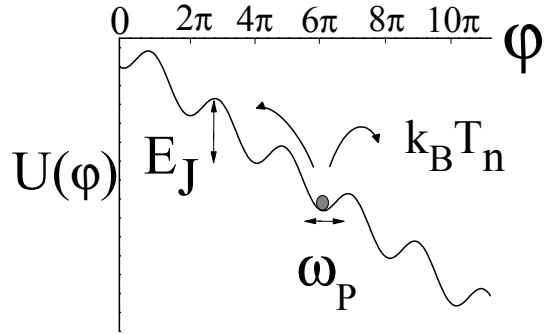

FIG. 1. Josephson phase dynamics of the washboard potential in the classical thermal fluctuation regime.

The phase diffusion model was first proposed by Ivanchenko and Zil'berman[24] and further developed by others.[25,26] In this model we can consider the thermal fluctuations as Johnson noise generated by a resistor $Z_{ENV}$ at a noise temperature $T_n$; both parameters depend only on the experimental set-up. In the limit of $\alpha = E_J/k_BT_n \leq 1$, they derived a simple analytic form for the *I-V* characteristics of the thermally fluctuated Josephson currents,

$$I(V) = \frac{I_C^2 Z_{ENV}}{2} \frac{V}{V^2 + V_P^2} \qquad (2)$$

As described above, the relevant energy scale of our STM Josephson junctions is $\alpha \leq 1$, so that the analytic form of the Eq. (2) is applicable to analyzing our data.

Now the pair current has a voltage dependence due to the diffusive phase motion. $V_P$, as a function of $T_n$ and $Z_{ENV}$, is the voltage where the pair current becomes maximum. $V_P$ will not change if $T_n$ and $Z_{ENV}$ are constant parameters intrinsic to the junction's environment, while $I_C$ increases as $R_N$ is decreased. Thus the thermally fluctuated Josephson current is characterized by three quantities, maximum supercurrent $I_C$, $T_n$ and $Z_{ENV}$. $T_n$ is an effective noise temperature for the ultra-small junction. This temperature can be elevated by noise from the room temperature electronics unless all the leads connecting to the junction are heavily filtered. $Z_{ENV}$ is the impedance of the junction's environment or the electronic circuit where the junction is embedded. It is reported that for ultra-small tunnel junctions, the Josephson phase dynamics is at very high frequency, characterized by the Josephson plasma frequency, $\omega_P$ or $E_J/\hbar$ (for STM Josephson junctions, it is of order of $10^{11} \sim 10^{12}$ Hz) and the frequency dependent damping at this frequency region is dominated by stray capacitance and inductance of the cables connecting the junction to the external circuit.[27,28,29,30] The cables will load the junction with an impedance on the order of the free space impedance, $Z_0 = \sqrt{\mu_0/\varepsilon_0} = 377\Omega$.

Experimentally we are interested in determining the Josephson $I_C R_N$ product, a quantity characteristic of the superconductivity of the constituent materials. If we observe the phase diffusion branches in our STM Josephson junctions, we can characterize them by identifying the two parameters, $T_n$ and $Z_{ENV}$ and then we can directly derive $I_C R_N$ of the material of interest by comparing the observed data and fits to the phase diffusion model.

Another concern for implementing a Josephson STM is the high current density due to the small geometry of the STM junction. A low junction resistance is desired such that $E_J$ is maximized, but it results in a high current density. In this configuration the current density, $j$, can be calculated using the tunnel current $I = 10$ nA which is a typical value for Pb/I/Pb STM Josephson junctions and the effective diameter of the superconducting tip, $\sim 3$ Å (for fcc structure of Ag, the nearest neighbor distance is 2.89 Å) over which electrons are being injected. This calculation yields $j \sim 10^7$ A/cm$^2$, a very high current density. Nevertheless as presented later, Josephson current was observed using the SC STM tip and no self heating effect due to the high current density was observed for Pb and NbSe$_2$.[31] For BSCCO, however, previous work reported that the high current density caused a huge effect on both its electronic structure and morphology.[5]

## IV. EXPERIMENTAL RESULTS

### A. Pb

To test the operation of our SC-STM, we first studied Ag-capped Pb films with our Ag-capped Pb tip (a symmetric junction). The spectrum we obtained is shown in Fig. 2 and is characteristic of S/I/S tunnel junctions: very sharp coherence peaks corresponding to $eV = \pm (\Delta_{tip} + \Delta_{sample})$, from which we obtained $\Delta_{tip} = \Delta_{sample} = 1.35$ meV, slightly smaller than the bulk value for Pb ($\Delta_{bulk} = 1.4$ meV) due to the proximity effect of the Ag capping layer. Moreover, the deviations from the BCS density of states outside the Pb gap due to strong-coupling effects are clearly seen at energies corresponding to the transverse and longitudinal phonon energies, $eV_T - 2\Delta$ = 4.5 meV and $eV_L - 2\Delta$ = 8.5 meV, respectively. Furthermore, just above the large coherence peaks we see the effects of the Ag proximity on the superconducting Pb. This shows up as a small dip just above the peak at $2\Delta$. Fig. 3 presents several I-V curves measured at different $R_N$ by changing the tip-sample distance sequentially. The superconducting gap size remains unchanged as $R_N$ is decreased. Low leakage current below the Pb gap (shown in Fig. 3) and the observation of the phonon structure in Fig.2 confirm high quality vacuum tunnel junctions. To acquire these I-V characteristics, the tunnel current feedback loop is temporarily turned off only during the measurement. We find that in the time required to do these measurements and the Josephson measurements later, the tip position remains stable. The feedback is re-established immediately after each I-V acquisition.

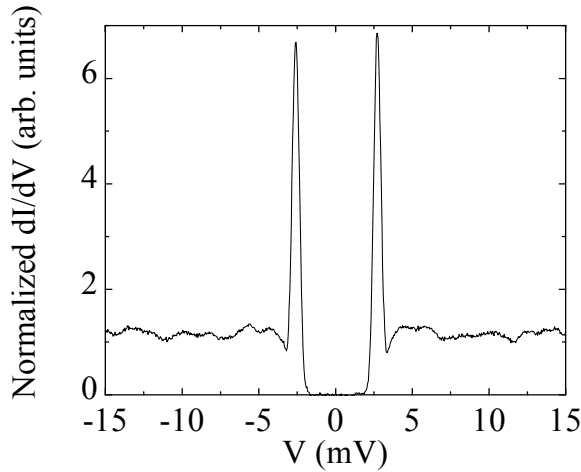

FIG. 2. Normalized $dI/dV$ spectrum of a Pb/I/Pb STM junction at $T$ = 2.1 K. The Pb phonon structures can be seen as dips at 7 and 12 mV: energies corresponding respectively to the transverse and longitudinal phonon energies in Pb as measured from the energy gap edge..

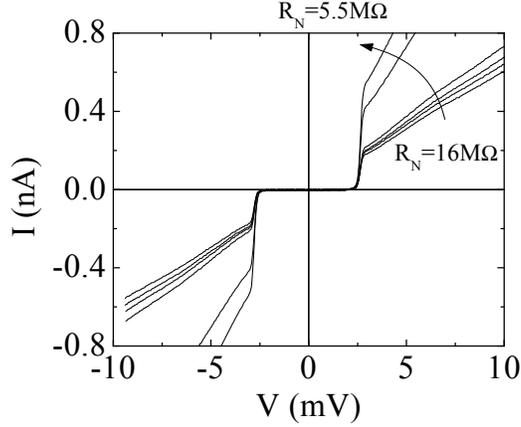

FIG. 3. *I-V* characteristics of Pb/I/Pb STM junctions at $T = 2.1$ K. *I-V* curves are measured as the junction normal state resistance $R_N$ is varied ($z$ position) with the tip position ($x,y$) fixed. At these high resistances, $E_J$ is too small for the Josephson effects to be observable. Also Coulomb blockade effects are negligible because the resistance of the tip and the film is such that the charge is swept away on a short time scale compared with $E_C$.

Fig. 4 shows the observed *I-V* characteristics at lower voltages (lines) for STM Josephson junctions formed between a Pb/Ag superconducting tip and a Pb/Ag superconducting film. The top panel of Fig. 4 shows that the location of the gap does not change as $R_N$ is lowered. The bottom panel is a close-up view of *I-V* characteristics near zero bias and clearly shows peaked structures first appearing and then exhibiting increasing heights as $R_N$ is decreased ($E_J$ enhanced). These observed *I-V* curves are fitted to the phase diffusion model (equation 2) with two parameters, $V_P$ and $I_C$.

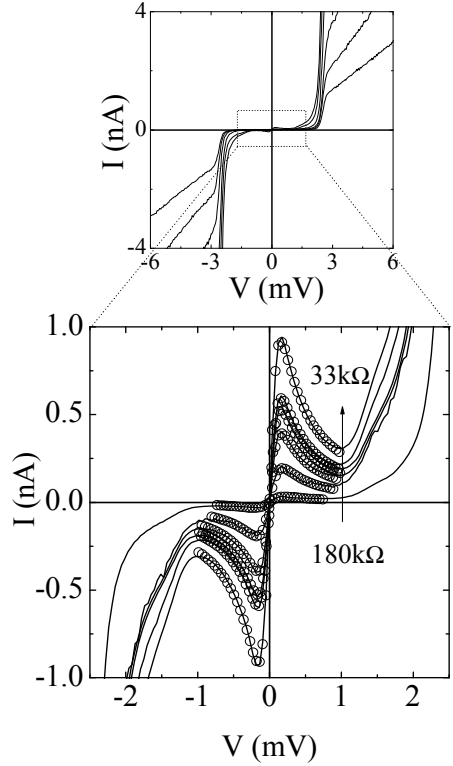

FIG. 4. *I-V* characteristics of Pb/I/Pb STM junctions at $T = 2.1$ K. (Top panel) Apparent are both the onset of the tunnel current at $V = 2\Delta$ and structures near zero bias (inside the box). (Bottom) *I-V* characteristics near zero bias for lower junction resistances than those in the top frame. The lines display the measured thermally fluctuated Josephson current and the symbols represent two-parameter fits to the phase diffusion model.

The best fits to the phase diffusion model are represented by the symbols in the bottom panel of Fig. 4 and the quality of these fits convince us that we have observed the signature of pair tunneling. This analysis yields a plot of $I_C \times \sqrt{e/k_B T_n}$ vs. $G_N = 1/R_N$, expected to be linear with zero intercept (no $I_C$ at infinite $R_N$) and a slope equal to $I_C R_N \times \sqrt{e/k_B T_n}$, as shown in Fig. 5(a). We can calculate $I_C R_N$ of Pb from the Ambegaokar-Baratoff formula[11] or Eq. (1) using $\Delta = 1.35$ meV at $T = 2.1$ K and including a factor of 0.788 due to strong electron-phonon coupling in Pb.[32] For $T = 2.1$ K and $T_C$(Pb) = 7.2 K, the hyperbolic tangent is very close to unity and we get $I_C R_N$ (Pb/I/Pb) = 1.671 mV. Substituting this value into the slope of the linear data fit in Fig 5(a) for our STM Josephson junctions, we can determine $T_n$ and $Z_{ENV}$, which are parameters depending only on the experimental set-up. Current values of these quantities for our apparatus are 15.9 ± 0.1 K and 279 ± 9 Ω, respectively. The fact that $T_n$ is higher than the 2.1K base temperature can be explained by leakage of rf noise to the junction from room temperature electronics. This value for $T_n$ is an improvement from our earlier work as a result of low temperature filters inserted close to the microscope. Further improvements are called for as $T_n$ is still higher than 2.1K. $Z_{ENV}$ is close to the expected value of the impedance of free space as described above. These quantities were measured using a preamplifier with $10^9$ gain. We repeated the Josephson measurements to derive $T_n$ and $Z_{ENV}$ using other preamplifiers with lower gains ($10^7$ and $10^8$) to cover the larger tunneling current range. All three data sets of the $I_C R_N$ plots in Fig. 5(b) fall on the same single line, convincing us that the Josephson coupling is enhanced as $R_N$ is decreased and $T_n$ and $Z_{ENV}$ are constant parameters intrinsic to the experimental circumstances and configuration. This also indicates that $T_n$ remains the same even for high current density (lower $R_N$). No heating effects or degraded superconducting DOS of the Pb tip or sample were observed

The lowest junction normal state resistance we studied was 4.6 kΩ, which is smaller than the quantum resistance of a single channel in the ballistic regime, $R_Q = h/2e^2 = 12.9$ kΩ. Sub-harmonic gap structures observed in the Pb/I/Pb STM Josephson junctions indicates that the low resistance STM junction is not in the weak tunneling limit ($|T|^2 = D = 10^{-6}$ or smaller), but somewhat higher transparency, $D \sim 0.1$ contributed from several conduction channels.[33]

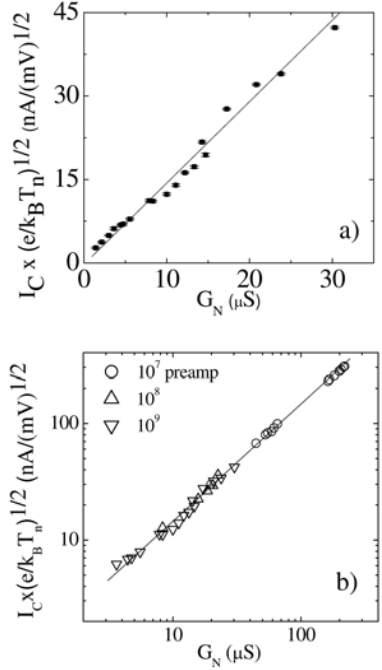

FIG. 5. (a) Plot of $I_C \times \sqrt{e/k_B T_n}$ vs. $G_N$ of Pb/I/Pb STM junctions. The slope is equal to $I_C R_N \times \sqrt{e/k_B T_n}$ and is shown as a linear fit. From the fitted slope and using the known value of $I_C R_N$ (Pb/I/Pb), $T_n$ and $Z_{ENV}$ are determined to be 15.9 ± 0.1 K and 279 ± 9 Ω, respectively. (b) Log-log plot of $I_C \times \sqrt{e/k_B T_n}$ vs. $G_N$ for data from three different preamplifiers (including that from 5(a). All data fall onto the same single line, indicating $T_n$ does not change due to large current flowing through the tunnel junction. These results give us confidence in using these two parameters $T_n$ and $Z_{ENV}$ later in our determination of $I_C R_N$ for NbSe$_2$ and BSCCO.

### B. NbSe$_2$

The data of Pb/I/NbSe$_2$ STM Josephson junctions are presented in Fig. 6. The experimental data on the bottom panel (lines) are contributions from the thermally fluctuated Josephson currents after subtraction of the quasiparticle background due to thermally excited quasiparticles. The background is obtained from the *I-V* curves of high resistance junctions where no Josephson currents are observed. Fig. 6 shows good agreement between the fits to the phase diffusion model (symbols) and the observed data. Since the slope value in Fig 7 is equal to $I_C R_N \times \sqrt{e/k_B T_n}$ and we can assume that $T_n$ and $Z_{ENV}$ remain constant, we can write a relationship $\sqrt{k_B T_n / e} = I_C R_N(Pb)/slope(Pb) = I_C R_N(Pb/NbSe_2)/slope(Pb/NbSe_2)$, where $slope(Pb)$ is obtained from the linear-fit to the data in the $I_C \times \sqrt{e/k_B T_n}$ vs. $G_N$ plot of the Pb/I/Pb STM Josephson junctions shown in Fig. 5. Substituting the known values, $I_C R_N(Pb)/slope(Pb)$ and the measured $slope$ (Pb/NbSe$_2$), we obtain $I_C R_N(Pb/NbSe_2) = 1.39 \pm 0.03$ mV. We then use the formula for the Josephson binding energy for different superconductors at $T = 0$ K given the gaps of each[34]

$$E_J = \frac{\hbar}{e^2 R_N} \frac{\Delta_1 \Delta_2}{\Delta_1 + \Delta_2} K\left(\frac{|\Delta_1 - \Delta_2|}{\Delta_1 + \Delta_2}\right) \tag{3}$$

where $K(x)$ is a complete elliptic integral of the first kind. Substituting $\Delta_1 = \Delta_{Pb} = 1.35$ meV and $\Delta_2$ for the smallest and average gap of NbSe$_2$, that is 0.7 and 1.1 meV respectively into equation 3 yields 1.34 mV $< I_C R_N$ (Pb/NbSe$_2$, $T = 0$ K) $< 1.70$ mV. Our result of 1.39 mV is in good agreement with the theoretical expectation.

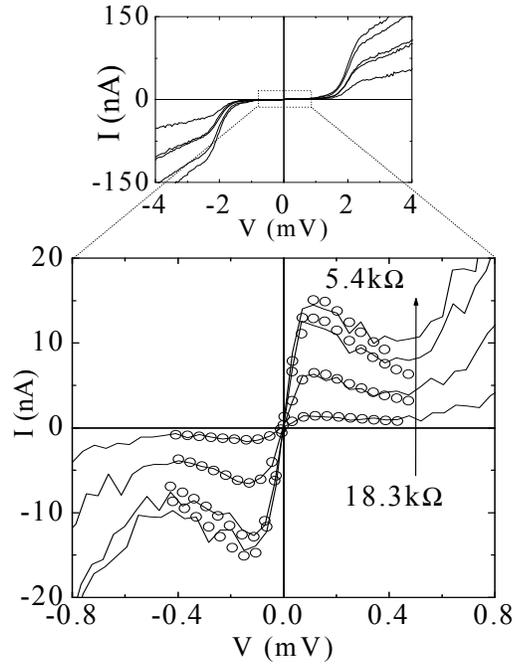

FIG. 6. *I-V* characteristics of Pb/I/NbSe$_2$ STM junctions at $T = 2.1$ K. (Top panel) Apparent is a current rise at V = $\Delta_{Pb}$ + $\Delta_{NbSe2}$. (Bottom) *I-V* characteristics near zero bias for lower junction resistances than those in the top frame, showing thermally fluctuated Josephson current. The symbols represent two-parameter fits to the phase diffusion model.

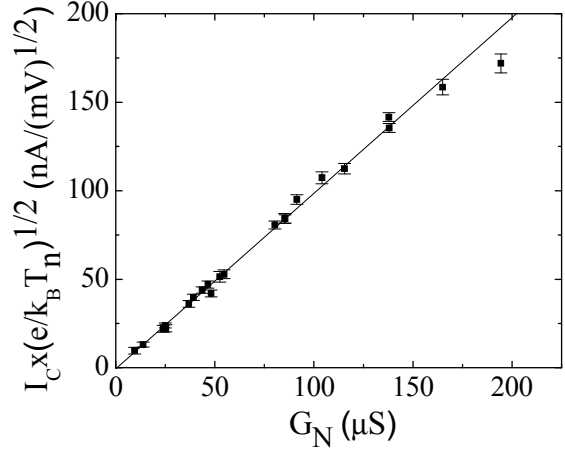

FIG. 7. Plot of $I_C \times \sqrt{e/k_B T_n}$ vs. $G_N$ for Pb/I/NbSe$_2$ STM junctions. The slope is equal to $I_C R_N \times \sqrt{e/k_B T_n}$ and is shown as a linear fit to the data. From the fitted slope and using the known value of $T_n$ = 15.9 K, $I_C R_N$ (Pb/NbSe$_2$) is determined.

### C. Bi$_2$Sr$_2$CaCu$_2$O$_{8+\delta}$

Fig. 8 is an atomic resolution image of cleaved optimally-doped BSCCO scanned by the superconducting STM tip at $T$ = 2.1 K. Because of the thick Pb layer used in our superconducting tip fabrication, it is difficult to routinely obtain atomic resolution images. However we can easily locate step edges and isolate flat surfaces where all the present data were obtained.

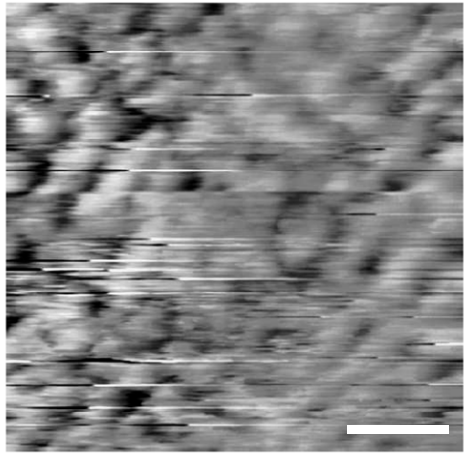

FIG. 8. Optimally-doped BSCCO topography scanned by superconducting STM tip at $T = 2.1$ K. The white bar is equal to 10 Å.

S/I/S STM junctions formed between the superconducting Pb tip and overdoped BSCCO single crystals show different features than those observed for the Pb/I/Pb STM junctions. Firstly, the energy gap of BSCCO is an order of magnitude larger than the Pb gap and secondly, the $dI/dV$ spectrum for the BSCCO gap has a "gaplessness" – non zero conductance at the Fermi energy – with an asymmetric normal state background conductance. Fig. 9(a) presents an I-V characteristic of Pb/I/overdoped BSCCO STM junctions at $T = 2.1$ K taken at $R_N = 10$ MΩ, clearly showing the Pb gap around 1.4 meV. The Pb gap edge does not have a sharp onset of tunnel current compared to that of Pb/I/Pb STM junctions because states exist all the way to the Fermi energy in the density of states of BSCCO. In other words, quasiparticles can tunnel at the Fermi energy of BSCCO. The inset of Fig. 9(a) shows a $dI/dV$ spectrum in the region of the Pb gap. The conductance outside the Pb gap is affected by the large energy gap of BSCCO ($\Delta_{BSCCO}$ = 40 meV as measured by the energy of the coherence peak.). Fig. 9(b) shows a $dI/dV$ spectrum taken with a large sweep range for the local density of states of BSCCO at the same location ($R_N$ = 500 MΩ).

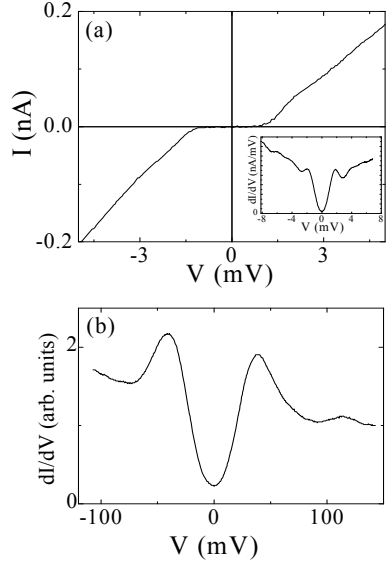

FIG. 9. (a) *I-V* characteristic of Pb/I/overdoped BSCCO STM junctions at $T$ = 2.1 K, clearly showing a Pb gap around 1.4 meV. Note the absence of leakage although the Pb gap edge is smeared compared to that of Pb/I/Pb STM junctions due to finite states all the way to the Fermi energy in the BSCCO density of states. Inset: $dI/dV$ for the Pb gap taken at $R_N$ = 10 MΩ. The conductance outside the Pb gap is affected by a relatively large energy gap of BSCCO ($\Delta_{BSCCO}$ = 40 meV). (b) $dI/dV$ spectrum taken over a large voltage range for the BSCCO gap at the same location. $R_N$ = 500 MΩ. The modulation amplitude added to the bias voltage is 2.5 mV$_{RMS}$.

For the local Josephson measurements for BSCCO single crystals, we first observe the $dI/dV$ spectrum at a particular surface point on overdoped BSCCO ($T_C$ = 79 K) in order to measure the energy gap Δ (solid line in the inset of Fig. 10). We use standard Lock-in techniques with 1 kHz modulation and a 2.5 mV$_{RMS}$ modulation voltage on the bias voltage and a junction normal resistance, $R_N$ ~ 500 MΩ. Although it is clearly a simplification of a more complex structure, we use the same definition for $\Delta$ as in previous works[35] in order to make comparisons. We then decrease $R_N$ to enhance $E_J$ in order to observe the pair tunnel current. A difference is that we cannot use very large currents with BSCCO due to the current limits for BSCCO damage (to be discussed below). This limitation also impacts the determination of $R_N$. Since the energy gap of BSCCO is much larger than that of Pb it is more difficult to measure $R_N$ from the *I-V* curves because of the limits on maximum current. Our procedure is to record several *I-V* curves by sweeping the bias voltage above the Pb gap. Then $R_N$ is determined by making use of our knowledge that the junction normal resistance inside the BSCCO gap (above the Pb gap) is 3 ~ 4 times larger than that outside the BSCCO gap determined from the $dI/dV$ spectrum such as illustrated in the inset of Fig. 10. For the lower junction resistance *I-V* curves, $R_N$ was calculated from the factor required to scale the current so that it overlaps with already normalized $I(V)R_N$ versus *V* curves with the ratio of the conductance inside to that outside the BSCCO gap. A set of $dI/dV$ data for the BSCCO gap and *I-V* curves taken with progressively lower junction resistances are necessary to calculate the $I_C R_N$ product every time the tip is moved to a new location.

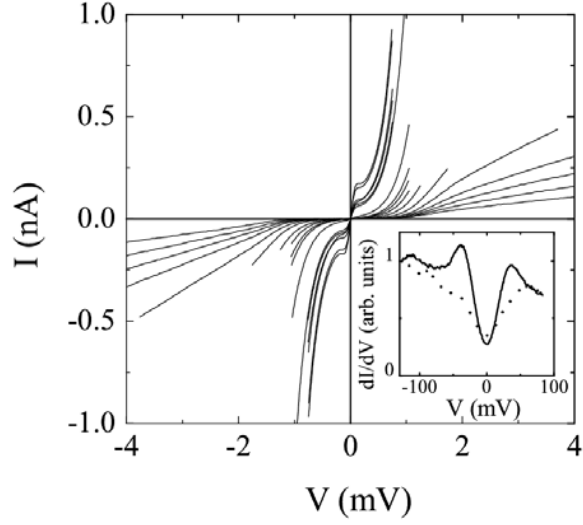

FIG. 10. *I-V* characteristics of Pb/I/overdoped BSCCO ($T_C$ = 79 K) STM Josephson junctions at $T$ = 2.1 K. The Pb gap is clearly seen around $V$ = 1.4 mV. Inset: $dI/dV$ spectrum (solid line) measured before low $R_N$ measurements, showing sharp coherence peaks with $\Delta$ = 37 meV. $dI/dV$ spectrum measured after low $R_N$ measurements (dotted line) indicates an LDOS change due to high current density.

In the main frame of Fig. 10 we plot the *I-V* characteristics at lower bias and lower $R_N$. A low leakage current below the Pb gap confirms the high quality of the vacuum tunnel junctions. Further decreasing $R_N$ increases the quasiparticle tunneling probability and finally the contribution from the thermally fluctuated Josephson currents is observed when $E_J$ is comparable to $k_B T_n$.[36]

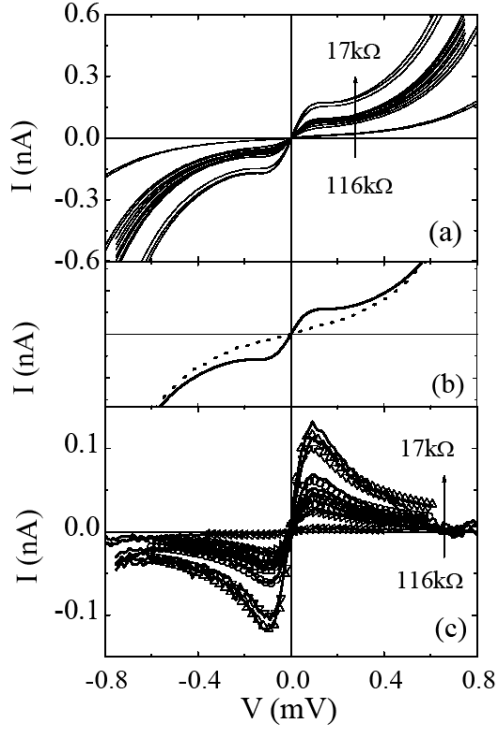

FIG. 11. (a) Low bias *I-V* characteristics of Fig. 10 for various junction resistances at $T = 2.1$ K. (b) Averaged *I-V* characteristic near zero bias for quasiparticle background (dotted line). One of the observed *I-V* curves is shown by the solid line. (c) Thermally fluctuated Josephson currents peaked at $V_P$ as derived by subtracting quasiparticle background (Fig. 11(b)) from the *I-V* curves (Fig. 11(a)). The data are represented by the lines and the symbols represent two-parameter fits to the phase diffusion model.

Fig. 11 displays a close-up view of the *I-V* characteristics near zero bias, clearly showing that the superconducting Pb tip was Josephson coupled to the BSCCO. The quasiparticle background represented by the dotted line in Fig. 11(b) is obtained from an average of several normalized *I-V* curves at higher $R_N$. At high $R_N$ there is no contribution from the Josephson effect and we use this curve as background. We scale it to the $R_N$ of the lower resistance data and the difference shown in Fig 11(b) is due to the Josephson currents. Fig. 11(c) shows the remaining contributions from the thermally fluctuated Josephson current after subtracting the quasiparticle background of Fig. 11(b) from the *I-V* curves of Fig. 11(a). The data in Fig. 11(c) are shown as lines and the best fits to the equation (2) are represented by the symbols. These good fits convince us that we have likely observed the pair current between a conventional (*s*-wave) superconducting Pb tip and overdoped BSCCO. This suggests that the BSCCO does not have a pure *d*-wave order parameter at least at the surface. In addition, the $dI/dV$ spectrum represented by the dotted line in the inset of Fig. 10 was observed after the lowest $R_N$ measurements in Fig. 11. The LDOS has changed significantly during the measurements; and the quasiparticle coherence peaks have disappeared, perhaps due to the high current density of the measurements at the highest conductance studied. This "modified" $dI/dV$ curve resembles those previously observed in heavily underdoped BSCCO,[7,9,10] in the "pseudogap" state at temperatures above $T_C$[7,37] and, in strongly disordered BSCCO thin films.[38] It is also similar to the $dI/dV$ spectra observed by others on surfaces which were altered by scanning with large tunnel currents.[5] It is important to note that LDOS changes were observed only after measurements were made with $R_N$ below 30 kΩ and *I* above the threshold current around 500 pA. Moreover the Josephson current disappeared after these irreversible changes of the LDOS on BSCCO occurred. In order to avoid this effect, most of the data presented here were obtained with $R_N$ ranging from 30 kΩ to 100 kΩ. This effect will be discussed in subsequent sections.

Each fit to the Josephson portion of the *I-V* curves in Fig. 11(c) generates a single data point in the plot shown in Fig. 12 in the similar way as described in the Pb/I/Pb and Pb/I/NbSe$_2$ STM junction results. As $G_N$ is increased ($R_N$ is reduced) the observed $I_C$ increases ($E_J$ increases). We now rely on the values for $T_n$ and $Z_{ENV}$ that we determined from our measurements on Pb/I/Pb for this experimental apparatus, and taking the slope of the linear fit shown in Fig. 12, we find $I_C R_N$ at this surface point to be 335 μV.

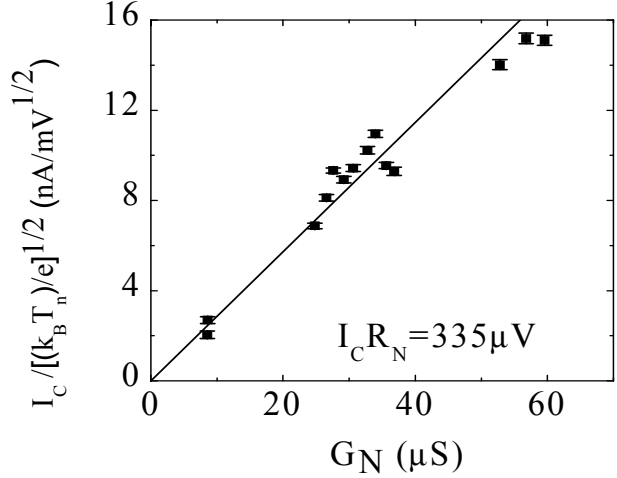

FIG. 12. Plot of $I_C \times \sqrt{e/k_B T_n}$ vs. $G_N$ of Pb/I/overdoped BSCCO ($T_C$ = 79 K) STM Josephson junctions. The slope is equal to $I_C R_N \times \sqrt{e/k_B T_n}$. Using the fitted slope and substituting the previously determined $T_n$, the Josephson product at this surface point is found to be $I_C R_N$ = 335 µV.

There are numerous normal tip STM studies of BSCCO aimed at developing a spatial picture of the electronic quantities of this material. Several of these investigations have produced renderings referred to as "gap maps".[35] These images reveal the inhomogeneous nature of the energy gap[6] and periodic electronic modulation both inside the vortex cores[39] and above $T_C$.[40] Again these data were derived from quasiparticle excitation spectra, not probing the superconducting pair state itself. It is natural to ask (i) whether the superconducting order parameter of BSCCO has spatial variation, and (ii) how the superconducting ground state correlates with the quasiparticle excited states ($\Delta$). Since we have the capability to measure both $\Delta$ (Fig. 10) and $I_C R_N$ (Fig. 11) at the same location on the surface, we have used these techniques to address these questions. In order to avoid the irreversible change in the LDOS for higher currents (Fig. 10 inset), the minimum junction resistance was kept above 30 k$\Omega$. The result for a second location on overdoped BSCCO ($T_C$ = 79 K) is presented in Fig. 13. Unlike the case shown in Fig. 10, the inset of Fig. 13(a) shows no appreciable change in the LDOS after the *I-V* measurement at the lowest $R_N$. We found $\Delta$ = 47 meV at this surface point. $I_C R_N$ derived from these data is also different from the previous location.

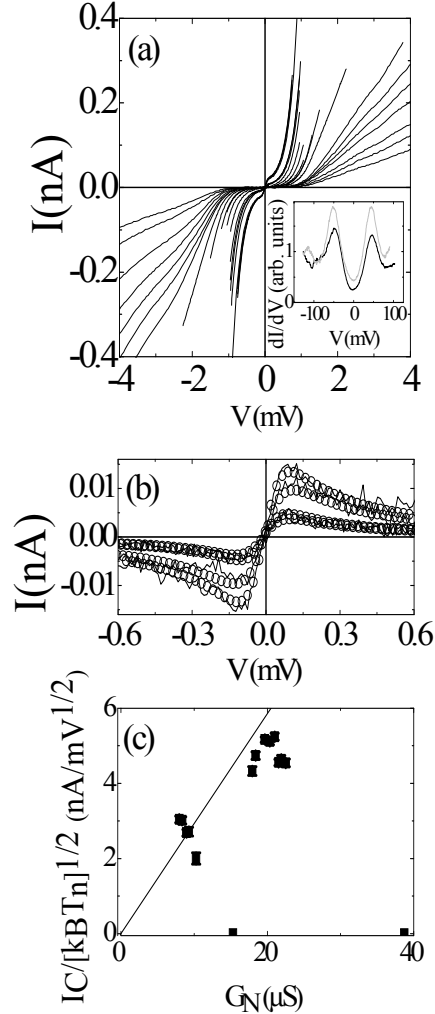

FIG. 13 *I-V* characteristics of Pb/I/overdoped BSCCO ($T_C$ = 79 K) STM Josephson junctions at different location from that in Fig. 10. (a) *I-V* characteristics of Pb/I/overdoped BSCCO ($T_C$ = 79 K) STM Josephson junctions at $T$ =2.1 K. Inset: $dI/dV$ measured before low $R_N$ measurement (black line) and after it (gray line). Note that energy gap is unchanged. (b) Thermally fluctuated Josephson currents (lines) and fits (circles) to the phase diffusion model. (c) Plot of $I_C \times \sqrt{e/k_B T_n}$ vs. $G_N$; $I_C R_N$ = 279 µV. Two data points at $G_N$ = 15 and 39 µS correspond to *I-V* characteristics without any pair current observed.

Of note in Fig. 13(c) are two data points showing zero $I_C$ appearing in the $I_C R_N$ plot. The disappearances of $I_C$ are observed at $G_N$ = 15 and 39 μS (*I-V* curves without any Josephson contributions). In order to check reproducibility of the pair current, we first repeated the Josephson current measurement at the same $R_N$ as that when currents had been observed. We also checked the linear relationship between $I_C$ and $G_N$ (both increasing and decreasing $G_N$). Reproducibility was usually observed. However, occasionally $I_C$ disappeared. Fig. 14 shows the $I_C R_N$ plot of Pb/I/overdoped BSCCO ($T_C$ = 76 K) measured at $T$ = 2.1 K. We first measured the *I-V* characteristic for the BSCCO gap (solid line in Fig. 14(a)). Then $R_N$ was decreased to see the Josephson current on the *I-V* characteristics. In Fig. 14(b), each Josephson measurement is labeled in the chronological order in which it was taken. The Josephson coupling increases as we decrease $R_N$ (expected) but then unexpectedly disappears at a lower $R_N$ (label 5). Increasing $R_N$ (decreasing $G_N$) results in the Josephson coupling returning (labels 6 and 7). This behavior is all unexpected. After these low $R_N$ measurements, the large bias *I-V* curve was again recorded to observe the BSCCO gap. This curve (dashes) in Fig. 14(a) indicates that the LDOS of the BSCCO and the energy gap remains the same before and after the low $R_N$ measurements so that the disappearance of $I_C$ was not caused by the LDOS change due to high current density. Although the origin of this disappearance is still under investigation, we observe that low $R_N$ measurements ($R_N$ below ~ 300 kΩ) on BSCCO, increases the low frequency noise on the tunnel current. This noise appears to be induced locally on the BSCCO and not from the environment or the electronics. These effects were not ever seen in our Josephson studies on the Pb-Pb or $NbSe_2$ systems. We assure ourselves that we have not affected the tip during the measurements by verifying that the Pb gap is always reproduced and the exponential decrease of the tunnel current vs. the tip-sample distance is also observed after low $R_N$ measurements.

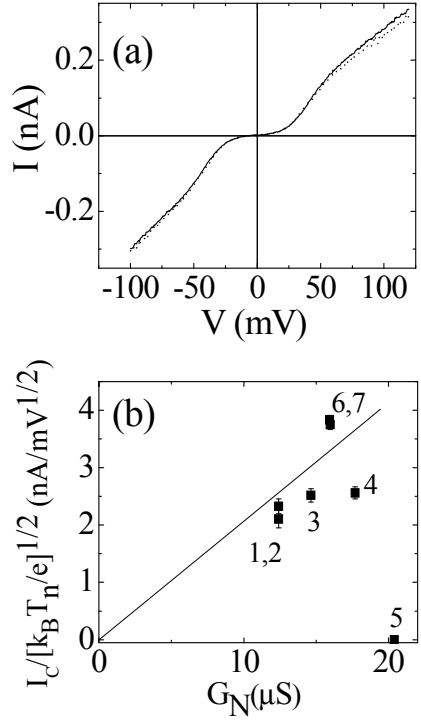

FIG. 14. *I-V* characteristics and $I_CR_N$ plot of Pb/I/overdoped BSCCO ($T_C$ = 76 K) STM Josephson junction for studying reproducibility of $I_C$. (a) *I-V* curve (solid line) measured before low $R_N$ measurements, showing $\Delta$ = 40 meV. *I-V* curve measured after low $R_N$ measurements (dotted line), indicating that the energy gap at this surface point remained the same. (b) Plot of $I_C \times \sqrt{e/k_BT_n}$ vs. $G_N$. The number labels represent the chronological order of the data sets. This order clearly shows the Josephson currents disappearing (5) and reappearing (6 and 7). Data points of (1,2) and (6,7) were measured repeatedly at $R_N$ = 80 and 63 kΩ, respectively.

Keeping these observations in mind, we performed both local Josephson and spectroscopic measurements on the overdoped BSCCO surface ($T_C$ = 79 K). Fig. 15 shows the spatial dependence of the energy gap and $I_CR_N$ measured simultaneously every 5 ~ 10 Å on a particular region on the surface. It clearly indicates that $I_CR_N$ and therefore the superconducting pair wave function of BSCCO changes on a nanometer-length scale on the surface. More interestingly, we can see an anticorrelation between $\Delta$ and $I_CR_N$ such that $I_CR_N$ tends to be reduced as $\Delta$ increases. This is not predicted by the BCS theory where $\Delta_{BCS}$ and $I_CR_N$ are linearly correlated. The tendency was also observed in the data taken along a line of 100 Å on another overdoped BSCCO ($T_C$ = 76 K) sample. This relationship is more apparent when $I_C R_N$ is plotted in the next section as a function of $\Delta$ for a variety of experiments.

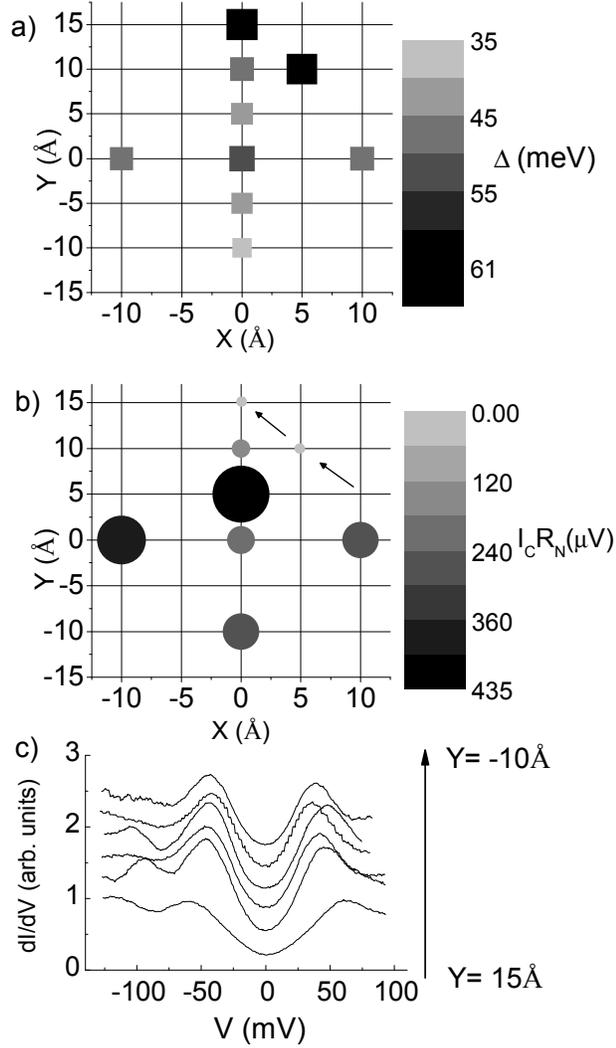

FIG. 15 Spatial studies of both (a) $\Delta$ and (b) $I_C R_N$ at the same locations on overdoped BSCCO ($T_C$ = 79 K). $I_C R_N$ changes spatially and seems to anticorrelate with $\Delta$. Note that no Josephson contributions were observed at the surface points denoted by arrows (in (b)) where the largest energy gaps were measured in this 20 Å × 25 Å region. (c) The line-cut of $dI/dV$ spectra are measured along the y-axis in (a), showing a well-known gap inhomogeneity (offset for clarity).

## V. DISCUSSION

### A. d-wave Superconductors

The observation of *c*-axis Josephson coupling in planar Pb -YBa$_2$Cu$_3$O$_{7-\delta}$ (YBCO) single crystal Josephson junctions has been reported and was explained by an *s*-wave component in the order parameter of YBCO induced by an orthorhombic distortion.[19] Although the crystallographic symmetry of Bi$_2$Sr$_2$CaCu$_2$O$_{8+\delta}$ (BSCCO) makes *s*- and *d*-wave mixing less likely,[41] Josephson coupling in planar junctions between conventional superconductors and BSCCO has been observed.[20,21] $I_CR_N$ values for these junctions (Nb- or Pb-BSCCO) ranged from 1 µV to 10 µV, suggesting that the *s*-component is about three orders of magnitude smaller than the *d*-component. Because $I_CR_N$ was measured in macroscopic junctions in previous work, any strong local inhomogeneities were obscured and meaningful comparisons with an inhomogeneous Δ could not be made. It is, therefore, very important to locally probe the order parameter in this strongly inhomogeneous material using Josephson tunneling.

For high-$T_C$ superconducting cuprates where the pairing mechanism is still under debate, attempts to extract the possible coupling due to the strong electron-phonon interaction were done for YBa$_2$Cu$_3$O$_{7-\delta}$ although the authors cautioned that the "gap" observed in the normalized conductance data for YBa$_2$Cu$_3$O$_{7-\delta}$ is not of the BCS form. Nevertheless, the expected $T_C$ was calculated from the normal state parameters, $\alpha^2F(\omega)$ and λ derived from the observed $dI/dV$ spectrum and found that it was 2/3 of the measured $T_C$ of this material.[42]

Recently microscopic studies of the phonon structure by STM were performed for $Bi_2Sr_2CaCu_2O_{8+\delta}$[43] and electron-doped cuprate, $Pr_{0.88}LaCe_{0.12}CuO_4$.[44] They have, however, extracted the phonon energies from positive peaks of $d^2I/dV^2$ spectra, with an assumption that the observed gap was equal to the superconducting gap, rather than following the previous procedure. Furthermore it has been suggested that their results could be interpreted as inelastic tunneling associating with apical oxygen within the barrier.[45,46] Electrons tunneling from the STM tip can lose energy to an oxygen vibrational phonon mode inside the barrier, yielding a new tunneling channel and mimicking the results reported. Furthermore, angle-resolved photoemission spectroscopy (ARPES) data has been interpreted within the context of a strong electron-phonon interaction model.[47]

## B. Variation of $I_CR_N$ in overdoped $Bi_2Sr_2CaCu_2O_{8+\delta}$

We interpret these results within the framework of the phase diagram for high-$T_C$ superconducting cuprates proposed by Emery and Kivelson.[48] High-$T_C$ superconducting cuprates are doped Mott insulators with low superfluid density, $n_S$. Therefore the phase stiffness, which is the energy scale to twist the phase, is small in these superconductors such that phase fluctuations could play an important role in determining $T_C$. There are two possible temperature scales that could affect the transition to superconductivity. $T_\theta$ is a temperature at which the phase ordering disappears because the phase stiffness disappears. Another temperature scale, $T^*$ is described as the temperature below which a gap on the quasiparticle spectrum appears. On the low doping side, a system could be divided into regions where the order parameter is well defined *locally* but not globally (a granular superconductor where the grains are weakly coupled to each other). These areas becomes larger with increasing doping, resulting in stronger inter-granular coupling so that the phase coherence length becomes longer and less susceptible to phase fluctuations. $T_\theta$ is increased as the hole doping, $\delta_h$ increases, leading to a rise of the global $T_C$ of the sample. Meantime, $T^*$ continues to decrease as $\delta_h$ increases. At the optimal doping, $T_\theta$ and $T^*$ cross over, so that the whole sample region is now phase coherent but the mean field value (energy gap) of the sample is suppressed. Thus $T_\theta$ and $T^*$ are the upper-bounds to $T_C$. $T_\theta$ is more significant due to the phase fluctuations on the lower doping side (underdoped), while $T^*$ is more important on the higher doping side (overdoped). Fig. 16 plots $T_C$ vs. hole doping, $\delta_h$ based on this Emery-Kivelson model and the superconducting region forms a dome shape with the maximum $T_C$ at $\delta_h \sim 0.16$ (optimally-doped). Decreasing or increasing $\delta_h$ from this value results in a $T_C$ decrease.

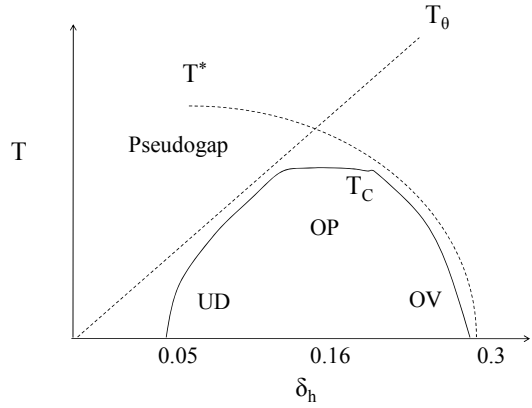

FIG. 16. Phase diagram based on the phase fluctuation model of high-$T_C$ superconductors as functions of temperature $T$ and hole doping, $\delta_h$ proposed by Emery and Kivelson. Optimally-doped (OP) region is a cross-over from underdoped (UD) to overdoped (OV) region. The phase ordering temperature, $T_\theta$, and the mean field transition temperature, $T^*$, are defined in the text.

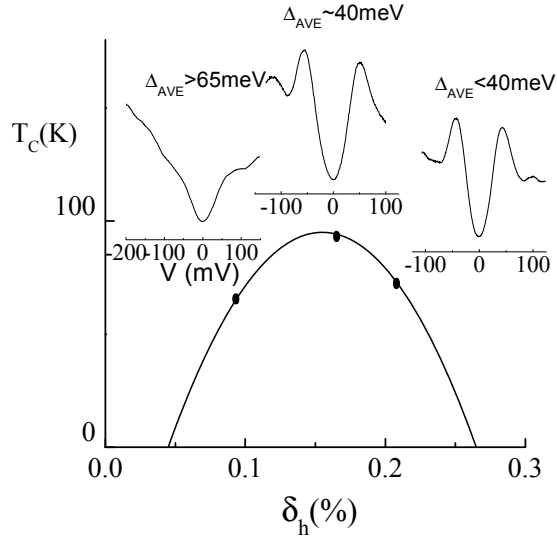

FIG. 17. Typical $dI/dV$ spectra and the corresponding averaged energy gaps, $\Delta_{AVE}$ at $T = 2.1$ K for BSCCO with three different dopings. They are underdoped ($T_C = 64$ K), optimally-doped ($T_C = 94$ K) and overdoped ($T_C = 76$ K) samples. Note that $\Delta_{AVE}$ monotonically decreases as $\delta_h$ increases.

In order to interpret our results using this Emery-Kivelson model ($T_C$ vs. $\delta_h$), we make two assumptions in order to replace $\delta_h$ in their model by the energy gap, $\Delta$ which we actually measure in our experiments.[36] First of all, $T^*$ changes monotonically with $\delta_h$ and decreases as $\delta_h$ increases. Previous STM studies[9,10] reported the spatially averaged gap value monotonically increased from overdoped (the average $\Delta \leq 40$ meV) to underdoped (the average $\Delta \geq 60$ meV), and $dI/dV$ with $\Delta \geq 65$ meV is often observed in heavily underdoped samples to no longer exhibit sharp coherence peaks. We also measured three samples with different dopings: underdoped ($T_C = 64$ K), optimally-doped ($T_C = 94$ K) and overdoped ($T_C = 76$ K) and observed this tendency as shown schematically in Fig. 17. The results indicate that the average $\Delta$, $\Delta_{AVE}$, seems to monotonically increase as $\delta_h$ is decreased. It was reported that the formation of gapped regions obtained from the $dI/dV$ spectra actually started above $T_C$ and there is a linear relation between $\Delta$ and the gap opening temperature, $T^*$ for optimally-doped and overdoped BSCCO samples.[7] Combining with all these facts, we suggest that the $\delta_h$-axis in the Emery-Kivelson model can be transformed into the $\Delta$-axis, but now $T^*$ monotonically increases with $\Delta_{AVE}$ as shown in Fig. 18.

Second, McElroy *et al.*[9] reported that all the gap-map studies for different dopings, ranging from underdoped to overdoped, show not only strong gap inhomogeneity over all samples, overdoped or underdoped (observation of the larger gap in regions of the overdoped and that of the smaller gap in regions of the underdoped samples), but also the shape of the averaged $dI/dV$ spectra for a given gap seems to be very similar, independent of whether the bulk sample is overdoped or underdoped. These lead us to make the second assumption that although the bulk (macroscopic) doping of each BSCCO sample is characterized by the transport $T_C$ and the spatially averaged energy gap, $\Delta_{AVE}$, local doping which will determine the local superconducting nature of the sample (the locally measured $\Delta$, the pair amplitude, $T_C$) reflects the observed inhomogeneity. Putting it another way, the smaller gap region which is sparsely distributed on the underdoped sample behaves as "overdoped", while the larger gap region which is rarely observed in overdoped sample behaves as "underdoped". This suggestion is also supported by the recent finding of local Fermi surface variations on BSCCO, indicating that local doping is not equivalent to the macroscopic doping of the sample.[49]

Since we measure $\Delta$, we choose to replot the Emery-Kivelson model schematically as shown in Fig. 18. We flip the $T_C$ vs. $\delta_h$ relation in the Emery-Kivelson model to a $T_C$ vs. $\Delta_{AVE}$ relation. Thus, the region, where the smaller gap is measured, we regard as an "overdoped" region, while the region with larger gap is regarded as "underdoped" region, even though our local Josephson measurements are done on overdoped samples ($T_C$ = 76, 79 K and 81 K). Now $T^*$ (and delta) monotonically increases and the dome-shaped region is simply flipped horizontally as shown when plotted vs. Δ.

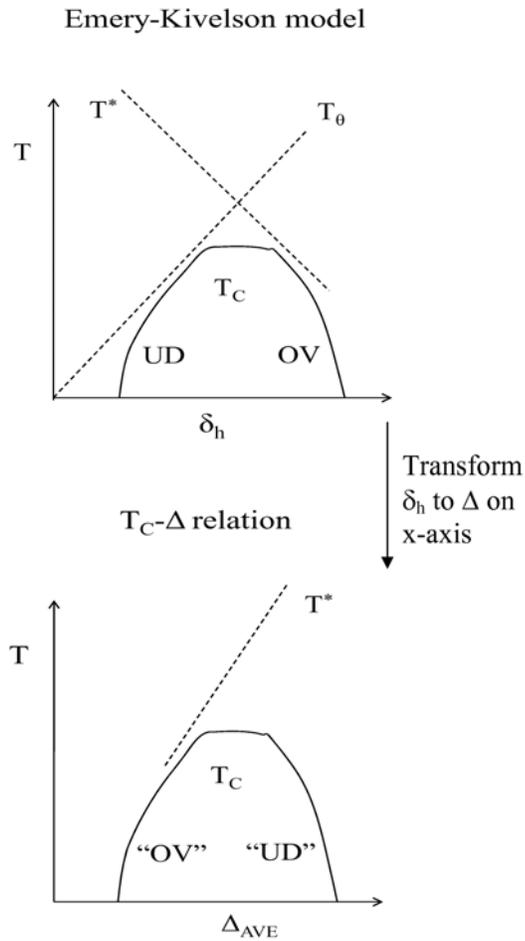

FIG. 18. Modified Emery-Kivelson model which includes two assumptions, (1) the linear relation between Δ and $T^*$ and, (2) a local doping variation on the BSCCO surface

With this adaptation in hand, we now summarize our measurements of the Josephson $I_C R_N$ product vs. $\Delta$ for five overdoped samples with each data point taken at locations roughly 5 ~ 10 Å apart.[36] Shown in Fig. 19 is the summary plot with the modified Emery-Kivelson model superimposed. Although there is scatter in the observed $I_C R_N$ values for a given $\Delta$, this figure clearly indicates the nanometer scale inhomogeneities in both $I_C R_N$ and $\Delta$. The reason for the scatter from experiment to experiment is under investigation. We believe this scatter is related to the microscopic inhomogeneities of BSCCO. We do not see this kind of scatter in the investigations of Pb/Ag or $NbSe_2$. A consistent but surprising feature seen from this plot is that $I_C R_N$ tends to be a maximum when $\Delta$ is between 40 and 45 meV, and the trend is for it to decrease or become zero as $\Delta$ increases or decreases from this maximal point.

Our results in Fig. 19 show that $I_C R_N$ is maximized at a gap value of 40 ~ 45 meV, the average $\Delta$ typically observed in optimally-doped BSCCO (corresponding to the highest $T_C$ samples). $I_C R_N$ decreases as $\Delta$ becomes larger. It also decreases as $\Delta$ becomes smaller. The $I_C R_N$ vs. $\Delta$ that we measure behaves in a similar way as $T_C$ vs. $\delta_h$ (as $\delta_h$ changes towards zero from the critical doping ~ 0.3 at the end of the superconducting region), and it does not follow the behavior of $\Delta$ vs. $T_C$ as expected from BCS theory. It is important to reiterate that for any given sample, we observed inhomogeneities both in $\Delta$ and $I_C R_N$ values as a function of location.

From our results we correlate the observed $I_C R_N$ with the amplitude of the superconducting order parameter $|\Psi|$ as well as with the $T_C$ of BSCCO via the Emery-Kivelson model phase diagram.[36] On the underdoped side of the phase diagram, these three quantities ($I_C R_N$, $|\Psi|$ and $T_C$) decrease (smaller superfluid density) as $\Delta$ increases and anticorrelate with $T^*$. This inverse relation between $I_C R_N$ and $\Delta$ in BSCCO is an unconventional result because in the BCS picture $\Delta_{BCS}$, $I_C R_N$, $|\Psi|$ and $T_C$ are all correlated. On the overdoped side of the phase diagram, $T_C$ decreases as $\delta_h$ is increased above 0.16 ($\Delta$ also becomes smaller in this doping region; a conventional result). Since the overdoped side is the amplitude dominated region, $T^*$ closely relates to $\Delta$ and hence decreases as $\delta_h$ is increased. $I_C R_N$ decreases as $\Delta$ is decreased from the value around 40 meV in Fig. 19, indicating $I_C R_N$, $|\Psi|$, $T_C$, $\Delta$ and $T^*$ behave similarly and conventionally as $\delta_h$ is increased towards the critical doping ($\delta_h \sim 0.3$) where $T_C$ vanishes.

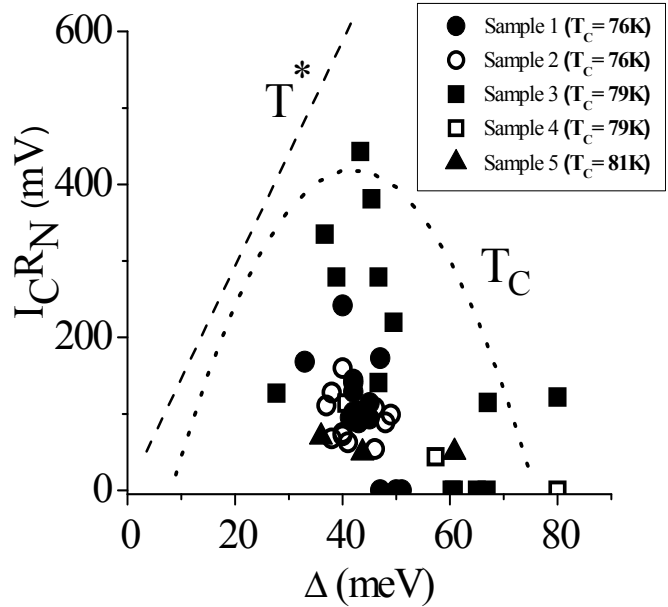

FIG. 19. $I_C R_N$ vs. $\Delta$ overlaid with the Emery-Kivelson model. Sketches of $T_C$ and $T^*$ from the Emery-Kivelson model are shown by dotted and dashed lines, respectively. The vertical scale for the model curves is arbitrary (obtained from ref. 36)

Another possible framework for discussing our results is the two-gap scenario observed in recent ARPES measurements.[50,51] In the underdoped regime in BSCCO samples, a smaller energy gap is observed and it becomes larger with increasing doping in the nodal region, distinct from the larger energy gap (pseudogap) in the antinodal region where no coherence peaks are observed. Moreover a temperature dependence of the nodal gap follows the BCS functional form very well, while the antinodal gap remains finite at $T_C$. The same trend of $\Delta(T)$ is also observed in overdoped samples, but the gapless region above $T_C$ expands on the Fermi surface with increasing doping (suppression of the pseudogap). It has also been reported that two gaps are observed in overdoped $(Bi_{1-y}Pb_y)_2Sr_2CuO_{6+x}$ using a variable temperature STM.[52] It was claimed that there are smaller, homogeneous energy gaps vanishing near $T_C$ as measured from the $dI/dV$ spectra normalized by normal state conductance, as well as the larger, inhomogeneous energy gaps, which have very weak temperature dependence. Although consistent with our $I_CR_N$ measurements, we do not observe the second gap *directly*. To our knowledge, the momentum $k$-component of the tunneling electron parallel to the junction barrier is conserved in the tunnel process, but the very small confinement of the electron due to the STM tip might increase the uncertainty of momentum and relax the constraint for momentum conservation. Thus the tunneling current observed in the STM could possibly be averaged over a large fraction of the momentum space, therefore, making it difficult to resolve a momentum dependent gap by STM. Moreover the results in Fig. 19 represent measurements of both $I_CR_N$ and $\Delta$ averaged over momentum space, and therefore we are unable to address this alternate model.

### C. Current density effect

We have observed notable changes in the LDOS after low $R_N$ (high current density) measurements. Most of these observations involve preliminary results, and more studies are necessary to come to a quantitative conclusion. Nevertheless, we have observed this effect so often that it deserves reporting. The effect is illustrated in the inset of Fig. 10. Several questions arise: Does the LDOS change suddenly or continuously? When or at what $R_N$ does it happen? How does LDOS change relate to the Josephson current? We performed "back and forth" measurements in which the BSCCO gap (high $R_N$ measurement) was measured followed by low $R_N$ measurements at the same location on the surface. Results are shown in Fig. 20. We first measured $dI/dV$ labeled 1, then measured I-V characteristics at lower voltages in the region of the Pb gap. $R_N$ is lowered until it reaches that of the I-V curve labeled 7. The tip is backed up to increase $R_N$ to measure $dI/dV$ for the BSCCO gap labeled 8 and so on. It is interesting that the BSCCO gap remains almost unchanged after measuring the Pb gap at $R_N = 68$ k$\Omega$, but a large LDOS change is observed ($dI/dV$ labeled 20) after obtaining the I-V curve labeled 19 at $R_N = 11$ k$\Omega$. The $dI/dV$ curve labeled 20 indicates not only a disappearance of sharp coherence peaks but an apparent increase in the energy gap size. Further decreasing $R_N$ to measure the Pb gap (I-V curve labeled 24) makes the LDOS change to a "V" shape ($dI/dV$ labeled 25) where we can no longer define an energy gap. From similar measurements, we have observed that the BSCCO gap and the shape of $dI/dV$ rarely change by measuring the Pb gap until $R_N$ is reduced to around 30 k$\Omega$, but further decreasing $R_N$ causes a deformation of LDOS. Howald *et al.* observed qualitatively similar $dI/dV$ curves on the intentionally disordered surface by scanning with large tunnel current.[5] Their tunnel condition, however, was at $I = 500$ pA with $V = -200$ mV so the power dissipated from the tip was $10^{-10}$ W, while the typical tunnel condition used in these current measurements for the I-V curve at $R_N = 30$ k$\Omega$, for example, is $I = 500$ pA with $V = 1.2$ mV so that the power dissipated from the superconducting tip is 100 times smaller than that used by Howald *et al.* although tip-sample distance in our STM junctions is smaller. In this configuration the current density, $j$, could be calculated using the tunnel current $I = 500$ pA and the effective diameter of the superconducting tip, $\sim 3$ Å over which electrons are being injected. Thus it yields $j \sim 10^6$ A/cm$^2$, a very high current density. It is still under investigation to answer why $R_N \sim 30$ kΩ is the threshold resistance for the BSCCO's LDOS change. We conclude that current density is the relevant parameter that causes these surface and spectral changes.

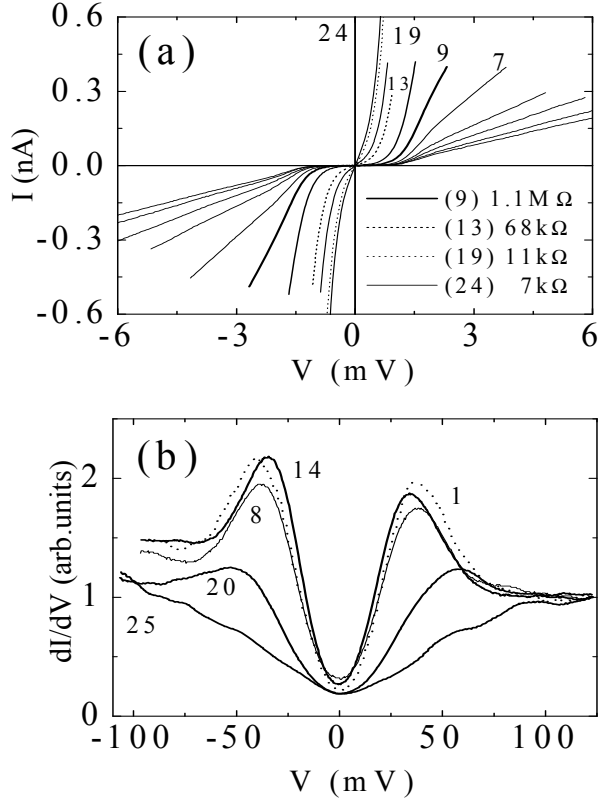

FIG. 20. (a) *I-V* characteristics and (b) $dI/dV$ spectra taken at the same surface point on overdoped BSCCO ($T_C$ = 74 K) at *T* = 2.1 K. LDOS change of this sample is caused by low $R_N$ measurements. The numbers labeled for *I-V* and $dI/dV$ curves are measured in chronological order. For example, $dI/dV$ spectrum (14) is taken right after *I-V* curve (13) was measured.

We have also measured the lateral range of the high current density alteration of the LDOS by moving the tip away from the original altered location to measure $dI/dV$ curves a few nanometers away. Fig. 21 shows the degraded LDOS is continuously changing, finally recovering to the superconducting LDOS as the tip is moved away from the originally damaged point. The $dI/dV$ spectrum with sharp coherence peaks is recovered at 13 nm away from the damaged point in opposite directions along a line. While Howald *et al.* "burned" the surface by scanning with a large tunnel current, we "burned" at a specific surface point by taking *I-V* characteristics at low $R_N$. It is interesting to note that spatial destruction of the BSCCO superconducting LDOS is similar for both experiments. Only qualitative studies of the high current density effect on BSCCO's electronic structure have been done so far, however, a relation between the thermally fluctuated Josephson current and LDOS change is still unknown. Further study is necessary to discuss this effect quantitatively.

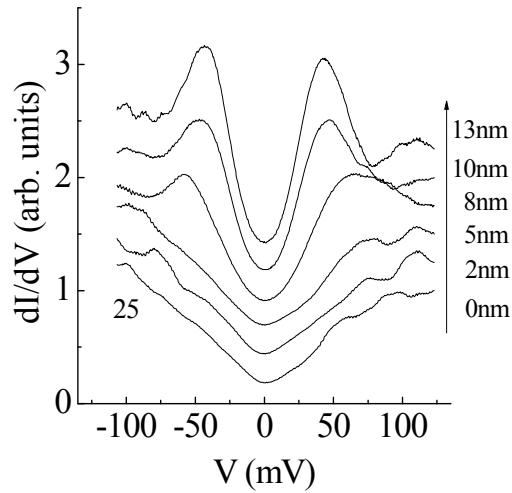

FIG. 21. $dI/dV$ spectra taken along a line from the originally damaged surface point (offset for clarity). The bottom curve is the same $dI/dV$ spectrum labeled 25 in Fig. 20(b).

### D.  Different surface preparation

Cleaving the BSCCO to expose an atomically flat surfaced is widely used for STM studies of this material; however, no study of the effect of the cleaving on the electronic structure of BSCCO has been reported. This fact results in a question as to whether the gap inhomogeneity routinely observed on cleaved surfaces is intrinsic to BSCCO or a result of the cleave. It is well known that the superconducting tunneling probes the depth of a coherence length into the sample surface. Therefore it is important to address this question because the electronic degradation on the surface of BSCCO could affect its tunneling current due to the very short *c*-axis coherence length ($\xi_C \leq 1$ nm) compared with much longer $\xi_C$ of conventional superconductors. Chemical etching is an alternate method to remove a degraded surface layer and possibly make a passivated layer. A chemical etching technique, originally reported by Vasquez *et al.*[53] was applied to Pb/I/YBCO tunnel junctions[54,55] and Josephson junctions.[19] An STM study of etched YBCO single crystals[56,57] revealed that etching with 1 % Bromine (Br) by volume in methanol resulted in an etching rate of 250 Å/min. The etching proceeds layer by layer and results in large flat areas separated by steps with single unit cell depth (~ 12 Å). The etching also produced pits on the surface which expand radially, introducing some surface roughness, but further etching removed layers without increasing roughness.

For BSCCO single crystals, we observed that 1 % Br in methanol was strong enough to dramatically roughen the etched surface, causing the STM to tip-crash. This result suggests that BSCCO is more sensitive to the etching than YBCO. In order to optimize the etching condition, etching rate calibrations were performed as follows: The BSCCO single crystal was coated with thinned rubber cement leaving a small region of exposed BSCCO. The etching solution consisting of 0.1 % Br in methanol was kept on the BSCCO single crystal for 3 minutes. The surface was then rinsed by dipping the sample in methanol in an ultrasound cleaner and the rubber cement was removed by sonication in toluene. A commercial profiler revealed clear steps at various edges of the etched region of roughly 6000 Å in height. This result indicates etch rates for BSCCO using 0.1 % Br of 2000 Å/min.

Fig. 22 shows a typical surface of overdoped BSCCO ($T_C$ = 74 K) etched in 0.1 % Br in methanol for 3 minutes followed by sonication in methanol and finally blown dry with nitrogen gas. The etched surface consists of "pancakes" with lateral dimensions ~ a few hundred Å. These pancakes have various step heights of not only a half unit cell in depth, 15 Å (Fig. 23), but also 5 Å and 10 Å. Fig. 24 shows a large area scan of the same sample as Fig. 22. A vertical corrugation over this surface is less than 30 Å, indicating that the 0.1 % Br etching proceeds layer by layer.

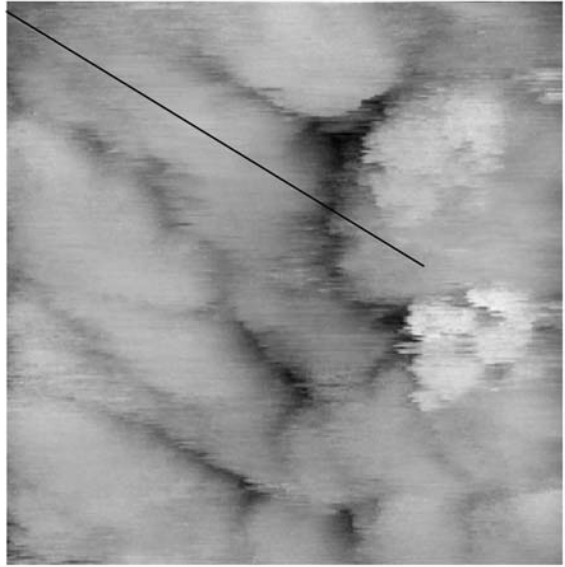

FIG. 22. STM image of 0.1 % Br etched overdoped BSCCO ($T_C$ = 74 K) at room temperature. Scan size is 800 Å × 800 Å.

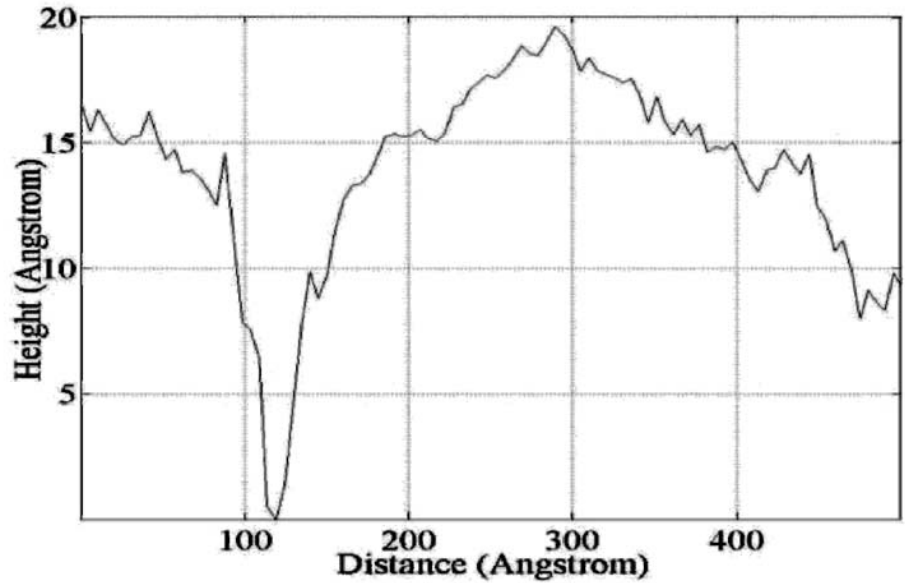

FIG. 23. Surface height cross section along the line shown in Fig. 22.

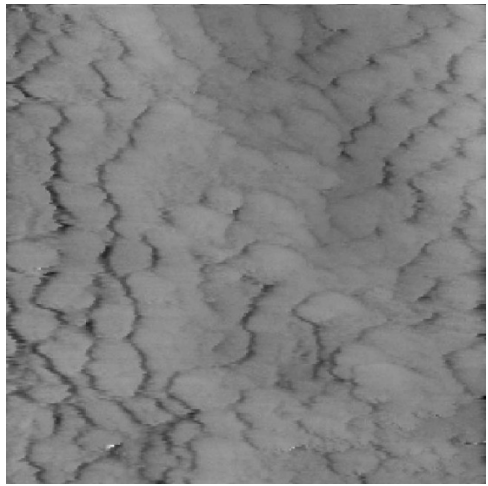

FIG. 24. Room temperature STM image of the same surface in Fig. 22 with larger scan area (3200 Å × 3200 Å).

Fig. 25 shows the $dI/dV$ spectra measured on the 0.1 % Br etched overdoped BSCCO sample at $T = 4.2$ K. Two $dI/dV$ curves were taken 10 Å apart. It is noteworthy that the spectral line shape looks very similar to that observed on the cleaved BSCCO sample except the asymmetry typically observed for cleaved samples in the coherence peaks is reversed. Nevertheless the result is reproducible. This sample was then cooled to $T = 2.1$ K to measure the Pb gap. Fig. 26 shows the *I-V* characteristic measured at a lower $R_N$. The Pb gap was clearly seen around $V = 1.4$ mV although the *I-V* curve was not taken at the same surface point as Fig. 26. Further investigation is required to determine whether the gap inhomogeneity is intrinsic to this material. It would also be useful to study how the $dI/dV$ spectra and the Josephson $I_C R_N$ product vary over an etched surface.

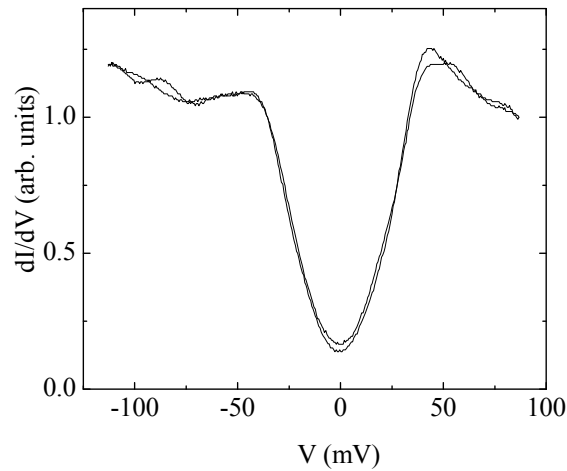

FIG. 25. $dI/dV$ spectra with a large bias measured on the 0.1 % Bromine etched overdoped BSCCO at $T = 4.2$ K. Two $dI/dV$ spectra were measured 10 Å apart.

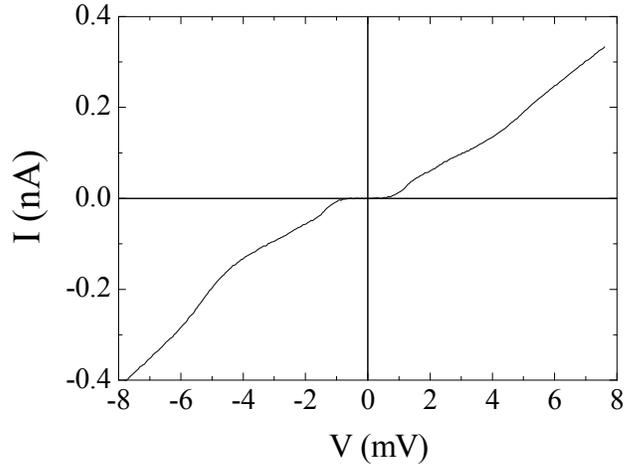

FIG. 26. *I-V* characteristic measured at lower $R_N$ on the 0.1 % Bromine etched overdoped BSCCO at $T$ = 2.1 K. The Pb gap is clearly seen around $V$ = 1.4 mV.

It is shown from these preliminary experiments that the Br etching proceeds layer by layer on BSCCO and yields a passivated surface. Observations of the superconducting Pb gap ensure that the passivated layer is thin enough for vacuum tunneling; however difficulty in reproducible observation of the Pb gap suggests that the thickness might change from etched sample to sample. Low temperature image scans also appear noisier than room temperature images, and there is difficulty at obtaining reproducible low temperature images. Since Br must be handled under a fume hood because of its high volatility and toxic nature, the etching process is done in air. This constraint possibly results in surface contamination although it still remains puzzling why the image is better (less noisy and reproducibly obtained) at room temperature than at low temperature.

In summary, we have prepared the BSCCO surface by chemical etching for our superconducting STM study. The etched BSCCO surface yields reproducible $dI/dV$ spectra, however extensive low $R_N$ measurements to observe Josephson current have not been accomplished yet.

## VI. CONCLUSION

We have described a series of experiments that utilize a superconductor-tipped STM to perform Josephson tunneling measurements on the high-$T_C$ superconducting cuprate, BSCCO. These measurements are motivated by the desire to directly access the pair wave function given the lack of a theory to connect the quasiparticle DOS to the superconducting state, in contrast to BCS superconductors. Operation of the apparatus has been verified by measurements of a conventional superconductor, Pb, and then the layered superconductor NbSe$_2$. We find good agreement between these measurements and theoretical predictions, giving us confidence in our BSCCO data for which no prediction is available. Our results indicate that like the quasiparticle DOS, the pair wave function is also inhomogeneous over the doping range studied, with variations on length scales of roughly 1 nm. Furthermore, we find that the gap measured from the quasiparticle DOS is anti-correlated with the Josephson $I_CR_N$ product for areas where the local superconducting nature has the characteristics and is consistent with underdoped samples. In addition we observe that excessive current densities can irreversibly alter the LDOS in these samples; and we have determined a limit on the current. Taking care to stay below those limits allows us to avoid this effect. In an effort to determine whether the local inhomogeneities are intrinsic or the result of the surface preparation by cleaving, we have also performed measurements of BSCCO samples that have been etched with a Br/methanol solution. Although no Josephson signal has been detected, this approach does appear to yield reasonable surfaces, albeit with apparent tip contamination challenges.

**ACKNOWLEDGEMENTS**

We thank the Berkeley Physics Machine shop for expert technical assistance and Garg Assoc Pvt Ltd for providing low noise miniature coax cable. The work in Berkeley was supported by DOE Grant No. DE-FG02-05ER46194. S.O. was supported by KAKENHI 20740213 and Y.A. by KAKENHI 19674002 and 20030004.